\documentclass[twocolumn,showpacs,preprintnumbers,amsmath,amssymb]{revtex4}

\usepackage{graphicx}
\usepackage{dcolumn}
\usepackage{bm}

\begin{document}
\draft

\unitlength 1mm

\title{Direct observation of a ``devil's staircase'' in wave-particle interaction}
\author{ Fabrice~Doveil}
\email{doveil@up.univ-mrs.fr}
\author{ Alessandro~Macor}
\email{macor@up.univ-mrs.fr}
\author{ Yves~Elskens}
\email{elskens@up.univ-mrs.fr}
\address{ Physique des interactions ioniques et mol{\'{e}}culaires,
                Unit{\'e} 6633 CNRS--Universit{\'e} de Provence, \\
                Equipe turbulence plasma,
                case 321, Centre de Saint-J{\'e}r{\^o}me,
                F-13397 Marseille cedex 20
                }

\begin{abstract}
We report the experimental observation of a ``devil's staircase''
in a time dependent system considered as a paradigm for the
transition to large scale chaos in the universality class of
hamiltonian systems. A test electron beam is used to observe its
non-self-consistent interaction with externally excited wave(s) in
a Travelling Wave Tube (TWT). A trochoidal energy analyzer records
the beam energy distribution at the output of the interaction
line. An arbitrary waveform generator is used to launch a
prescribed spectrum of waves along the slow wave structure (a 4 m
long helix) of the TWT. The resonant velocity domain associated to
a single wave is observed, as well as the transition to large
scale chaos when the resonant domains of two waves and their
secondary resonances overlap. This transition exhibits a ``devil's
staircase'' behavior for increasing excitation amplitude, due to
the nonlinear forcing by the second wave on the pendulum-like
motion of a charged particle in one electrostatic wave.
\newline %
\end{abstract}

\pacs{05.45.Df, 52.40.Mj, 52.20.Dq, 84.40.Fe}

\maketitle

%



{\bf Numerous nonlinear driven systems in physics, astronomy, and
engineering exhibit striking responses with complex phase-locking
plateaus characterized by devil's staircases. Whereas this
structure has fascinated theoreticians since its discovery, and
was observed in various condensed matter systems, no dynamical
system had so far allowed its direct observation. We realized such
an observation for the motion of a charged particle in the
potential of two waves. This system has already been used as a
theoretical and experimental paradigm for analysing the transition
to large scale chaos in conservative systems. Our method consists
in measuring the velocity distribution function of a test beam at
the outlet of a harmonically excited travelling wave tube. A
devil's staircase is directly measured by varying the amplitude of
the excitation in perfect agreement with numerical simulations.
This clear experimental realization for a theoretical paradigm
opens the possibility to achieve the detailed exploration of
chaotic behaviour in non-dissipative systems.}

\section{Introduction}

Wave-particle interaction is central to the operation of electron
devices and to the understanding of basic plasma behavior and more
generally of systems with long range interaction or global
coupling. Specifically, travelling wave tubes are ideal to study
wave-particle interaction \cite{Dim,Tsu,APS05}. In the experiments
reported here, the electron beam density is so low that it does
not induce any significant growth of the waves. When waves are
launched in the tube with an external antenna, the beam electrons
can be considered as test particles in the potential of
electrostatic waves. This recently allowed the direct exploration
of nonlinear particle synchronization \cite{DovEsc} by a single
non resonant wave with a phase velocity different from the beam
velocity.

In this paper we focus on the resonant case when the beam and the
waves propagate with similar velocities. For a single wave, we
observe beam trapping in the potential troughs of the wave. When
at least two waves propagate we observe a large velocity spread of
the essentially mono-kinetic initial beam. This spread can be
related to a transition to large scale chaos through resonance
overlap and destruction of invariant KAM (Kolmogorov-Arnold-Moser)
tori \cite{DovAu}. We also observe that the velocity spreading is
not smooth and occurs through successive steps as the waves
amplitudes increase.

The motion of a charged particle in two electrostatic waves has
been used as a paradigm system to describe the transition to
large-scale chaos in Hamiltonian dynamics
\cite{Chi,Esc,Cha,Ko,ElsEsc}. It is well known that Hamiltonian
phase space exhibits a scale invariance with infinitely nested
resonant islands \cite{DovEsca}. We show that the observed steps
in the measured velocity distribution function are related to the
underlying ``devil's staircase'' behavior.

Numerous nonlinear driven systems in physics, astronomy, and
engineering exhibit striking responses with complex phase-locking
plateaus characterized by devil's staircase, Arnold tongues, and
Farey trees \cite{Lic,Kad,Sch,Ott}. These structures are mainly
seen at work in stationary and space-extended objects, when an
elastic lattice interacts with a periodic substrate found in a
wide variety of condensed matter systems including charge-density
waves \cite{Gor}, Josephson junction arrays \cite{Ori},
Frenkel-Kontorova-type models of friction \cite{Flo}, unusual
growth mode of crystals \cite{Hup}, and in semiconductor-metal
phase transition \cite{Ade}. They are also at work in the quantum
Hall effects \cite{Das}. But very few dynamical systems have so
far allowed their observation. Only indirect observations of
statistically locked-in transport through periodic potential
landscapes have been made \cite{Gop}. Here we present a direct
experimental evidence of a ``devil's staircase'' for wave-particle
interaction \cite{Mac}.

The paper is structured as follows: In Sec.~II we present the
principle of the test beam experiment in the TWT. In Sec.~III, we
report the main results on the interaction of the test beam with
one or two waves. In Sec.~IV, we briefly recall the main results
on the transition to large scale chaos in Hamiltonian systems and
show some numerical results for the paradigm motion of a charged
particle in two waves. In Sec.~V, we compare the experimental
results with the numerical predictions. Conclusions and
perspectives are drawn in Sec.~VI.

\section{Description of the TWT}

A travelling wave tube \cite{Pie,Gil} consists of two main
elements: an electron gun, and a slow wave structure along which
waves propagate with a phase velocity much lower than the velocity
of light, which may be approximately equal to the electron beam
velocity.

The long travelling wave tube sketched in Fig.~1 has been designed
to mimic beam-plasma interaction \cite{Dim,Tsu,APS05}. Its slow
wave structure is formed by a helix with axially movable antennas.
The electron gun produces a quasi-mono-kinetic electron beam which
propagates along the axis of the slow wave structure and is
confined by a strong axial magnetic field of $0.05$~T. Its central
part consists of the grid-cathode subassembly of a ceramic
microwave triode and the anode is replaced by a copper plate with
an on-axis hole whose aperture defines the beam diameter equal to
$1 {\,\rm mm}$. Beam currents, $I_{\mathrm b} < 1 {\,\rm {\mu
A}}$, and maximal cathode voltages, $|V_{\mathrm c}| < 200 {\,\rm
V}$, can be set independently. Two correction coils (not shown in
Fig.~1) provide perpendicular magnetic fields to control the tilt
of the electron beam with respect to the axis of the helix.

The slow wave structure consists of a wire helix that is rigidly
held together by three threaded alumina rods and is enclosed by a
glass vacuum tube. The pump pressure at the ion pumps on both ends
of the device is $2 \times10^{-9} {\,\rm Torr}$. The $4 {\,\rm m}$
long helix is made of a $0.3 {\,\rm mm}$ diameter Be-Cu wire; its
radius is equal to $11.3 {\,\rm mm}$ and its pitch to $0.8 {\,\rm
mm}$. A resistive rf termination at each end of the helix reduces
reflections. The maximal voltage standing wave ratio is $1.2$ due
to residual end reflections and irregularities of the helix. The
glass vacuum jacket is enclosed by an axially slotted $57.5 {\,\rm
mm}$ radius cylinder that defines the rf ground. Inside this
cylinder but outside the vacuum jacket are four axially movable
antennas labelled from ${\sharp\,1}$ to ${\sharp\,4}$ and
azimuthally separated by $90\,^{0}$; they are capacitively coupled
to the helix, each of them being characterized by its coupling
coefficient which gives the ratio of the amplitude of the wave
actually excited on the helix to the amplitude of the signal
launched on the antenna; these antennas can excite or detect helix
modes in the frequency range from 5 to 95 {\,\rm MHz}. Only the
helix modes are launched, since empty waveguide modes can only
propagate above $2 {\,\rm GHz}$. These modes have electric field
components along the helix axis \cite{Dim}. Launched
electromagnetic waves travel along the helix at the speed of
light; their phase velocities, ${v_{{\mathrm \varphi} j}}$, along
the axis of the helix are smaller by approximately the tangent of
the pitch angle, giving $2.8 \times10^6 {\,\rm m/s} < {v_{{\mathrm
\varphi} j}} < 5.3 \times10^6 {\,\rm m/s}$. A beam intensity
$I_{\mathrm b}$ lower than $150$~nA ensures that the electron beam
with mean velocity $v_{\mathrm b}$ does not induces any
significant wave growth over the tube length; since the potential
drop over the beam radius $\delta V = I_{\mathrm b}/(4 \pi
\epsilon_0 v_{\mathrm b})$ is negligible, the beam electrons can
be considered as test particles.

Finally the cumulative changes of the electron beam distribution
are measured with a trochoidal velocity analyzer at the end of the
interaction region. A small fraction ($0.5\%$) of the electrons
passes through a hole in the center of the front collector, and is
slowed down by three retarding electrodes. By operating a
selection of electrons through the use of the drift velocity
caused by an electric field perpendicular to the magnetic field,
the direct measurement of the current collected behind a tiny
off-axis hole gives the time averaged beam axial energy
distribution \cite{Guy}. Retarding potential and measured current
are computer controlled, allowing an easy acquisition and
treatment with an energy resolution lower than $0.5 {\,\rm eV}$.

\section{Test particle experiments}
\label{SecExp}

We first checked that, in the absence of any applied external
signal on the antenna, the test electron beam propagates without
perturbation along the axis of the device.

\subsection{Trapping of the test beam in a resonant wave}
In the first experiment, we apply an oscillating signal at a
frequency of $f = 30$~MHz on antenna ${\sharp\,1}$. According to
the helix dispersion relation, a travelling wave propagates along
the helix with a phase velocity $v_\varphi  = 4.07
\times10^6$~m/s. Fig.~\ref{fig2}a shows a 2D contour plot (with
logarithmically scaled color coding) of the velocity distribution
function for a test beam, with intensity $I_{\mathrm b} = 120$~nA
and initial velocity $v_{\mathrm b} = v_\varphi$, measured at the
outlet of the tube after its interaction, over a length of
$2.6$~m, with the helix mode. It is the result of linear
interpolation of the recorded measurements obtained for different
applied signal amplitudes varying from $0 {\,\rm mV}$ to $1400
{\,\rm mV}$ by steps of $100 {\,\rm mV}$. Each velocity
distribution function measurement is obtained by scanning the
retarding voltage of the trochoidal velocity analyzer with a step
of $61$~mV. When the wave amplitude is gradually increased, we
observe that the width of the velocity domain in which the test
beam electrons are spread increases like the square root of the
wave amplitude as expected if the electrons are trapped in the
potential troughs of the wave. No attention must be given to the
apparent regular discontinuous steps due to bad sampling of the
amplitudes and interpolation between successive measurements. From
the measured antenna coupling coefficients \cite{Malmberg}, the
helix wave amplitude $\phi$ can be estimated. The resonant or
trapping domain in velocity can thus be deduced and is indicated
by the continuous lines in Fig.~\ref{fig2}a which correspond to
$v_{\varphi} \pm 2\sqrt{\eta \phi}$; here $\eta$ is the charge to
mass ratio of the electron. We observe a very good agreement with
measurement.

Fig.~\ref{fig2}b is obtained in the same way as Fig.~\ref{fig2}a
but corresponds to a larger interaction length since the emitting
antenna is located at $L = 3.5$~m from the device output. Each
velocity distribution function measurement is obtained by scanning
the retarding voltage of the trochoidal velocity analyzer with a
step of $240$~mV to speed up data acquisition without
deteriorating significantly the accuracy of our observations. We
observe that the shape of the velocity domain in which the test
beam electrons are spread is very different from Fig.~\ref{fig2}a.
A new feature appearing in Fig.~\ref{fig2}b is a further velocity
bunching of the electrons around their initial velocity for an
amplitude equal to $600 {\,\rm mV}$. This phenomenon is also
related to the trapping of the electrons in the wave. If we refer
to the rotating bar model \cite{Myn} to describe the trapped
electrons motion, we expect oscillations between small spread in
velocity and large spread in position (corresponding to initial
conditions for a cold test beam) to large spread in velocity and
small spread in position after half a bounce or trapping period
equal to $T_b /2 = v_{\varphi} /(2 f \sqrt{\eta \phi})$. To the
bounce period, we can associate a bounce length $L_b = v_\varphi
T_b$. For $\phi = 35 {\,\rm mV}$, we get $L_b/2 = 350 {\,\rm cm}$
which is precisely the interaction length $L$ of Fig.~\ref{fig2}b.
We thus confirm that Fig.~\ref{fig2} is displaying the trapping of
the test beam in a single wave. Varying $I_{\mathrm b}$ from $10$
to $150$~nA would not change the result since no self-consistent
effect occurs. One may remark that Fig.~\ref{fig2} also displays a
systematic accumulation of particles in the high velocity region
for which we do not have a simple explanation and which probably
requires further investigation out of the scope of the present
paper.

\subsection{Test beam interacting with two waves}
In the second experiment, we apply, on antenna ${\sharp\,2}$
chosen for its strongest coupling with the helix and located at $L
= 3.5$~m from the device output, a signal generated by an
arbitrary waveform generator made of two components with well
defined phases: again one at the same frequency of $30$~MHz, and a
new one at a frequency of $60$~MHz. According to the helix
dispersion relation, two travelling waves propagate along the
helix, the former with a phase velocity $v_{\varphi 1}=
4.07\cdot10^6$~m/s as before, the latter with a different phase
velocity $v_{\varphi 2}= 3.08\cdot10^6$~m/s (the fact that the
second wave is a harmonic of the first one in the laboratory frame
is irrelevant in the beam frame).

Fig.~\ref{fig3}a (resp. Fig.~\ref{fig3}b) is obtained in the same
way as Fig.~\ref{fig2} for a test beam, with intensity $I_{\mathrm
b} = 10$~nA and initial velocity $v_{\mathrm b} = v_{\varphi 1}$
(resp. $v_{\mathrm b} = v_{\varphi 2}$), whose velocity
distribution function is measured at the outlet of the tube after
its interaction with the two propagating helix modes. As in
Fig.~\ref{fig2} the continuous (resp.~dashed) parabola shows the
trapping velocity domain associated to the helix mode at $60$~MHz
(resp.~$30$~MHz). As it appears clearly, the amplitudes of the two
modes have been appropriately chosen such that these trapping
domains approximately have the same velocity extension when taking
into account the coupling coefficient of the antenna at the two
working frequencies. Both in Fig.~\ref{fig3}a and
Fig.~\ref{fig3}b, we observe that, for a threshold amplitude
exceeding the amplitude corresponding to overlap of the trapping
domains, the velocity distribution function exhibits a strong
velocity spread. Electrons initially trapped inside the potential
well of one of the helix modes can escape the trapping domain and
explore a much wider velocity region in phase space. According to
their initial velocity, electrons can be strongly accelerated
(Fig.~\ref{fig3}a) or decelerated (Fig.~\ref{fig3}b) sticking
mainly to the border of the new velocity domain they are allowed
to explore. We shall later relate this behavior to the transition
to large scale chaos for the motion of a charged particle in two
electrostatic waves. Such a description would suggest that
Fig.~\ref{fig3}b could be obtained by a mere reflection of
Fig.~\ref{fig3}a with respect to the average velocity of the two
waves. The observed asymmetry can result from the different
transit times of the test particles over the finite length of the
tube in both cases, and from the already noticed accumulation of
particles toward higher velocities of Fig.~\ref{fig2}.

\subsection{Test beam and excitation at a single frequency}
In the third experiment, we apply an oscillating signal at a
frequency of $30$~MHz on antenna ${\sharp\,2}$ as in the first
experiment for maximum interaction length $L = 3.5$~m. But we now
consider a test beam, with intensity $I_{\mathrm b} = 10$~nA and
initial velocity $v_{\mathrm b} = 2.7\cdot10^6$~m/s much lower
than $v_\varphi$, the phase velocity of the helix mode at
$30$~MHz. Fig.~\ref{fig4} is obtained in the same way as
Fig.~\ref{fig2}. The continuous parabola indicates the velocity
domain associated to particle trapping in the potential troughs of
the helix mode at $30$~MHz. Since the initial beam velocity lies
far out of this domain we expect that, for moderate wave
amplitude, the beam electrons will experience a mere velocity
modulation around their initial velocity, with a modulation
amplitude increasing linearly with the applied signal amplitude.
This behavior has been studied in detail in Ref.~[3] and would
generate two main peaks around the (oblique straight) continuous
lines originating in $v_{\mathrm b}$.

In fact we observe that the electrons spread over a velocity
domain with typical width increasing as the square root of the
applied signal amplitude as shown by the dashed parabola. This
strongly recalls the results of Fig.~\ref{fig2}. Indeed the
applied signal generates two waves: a helix mode with a phase
velocity $v_\varphi$, and a beam mode with a phase velocity equal
to the beam velocity $v_{\mathrm b}$. The beam mode with phase
velocity $v_{\mathrm b}$ is actually the superposition of two
indistinguishable modes with pulsation $\omega= kv_{\mathrm b} \pm
\omega_{\mathrm b}$ corresponding to the beam plasma mode with
pulsation $ \omega_{\mathrm b} = [n_{\mathrm b} e^2/(m
\epsilon_0)]^{1/2}$ for a beam with density $n_{\mathrm b}$,
Doppler-shifted by the beam velocity $v_{\mathrm b}$, merging in a
single mode since $ \omega_{\mathrm b} \ll \omega $  in our
conditions \cite{Pie}. Thus Fig.~\ref{fig4} shows the test
electrons trapping into the beam mode. This is confirmed by a
careful analysis of Fig.~\ref{fig4} which exhibits the same
velocity bunching of the electrons around their initial velocity
as in Fig.~\ref{fig2}a, for amplitudes obtained by equating the
interaction length to a multiple of half the trapping length. One
also notices that the amplitude of the beam mode is lower than the
amplitude of the launched helix mode; this explains why its
influence has been neglected in the previous analysis of
Fig.~\ref{fig3} where two helix modes are externally excited.
Since, when we increase the applied signal amplitude in
Fig.~\ref{fig4}, the two trapping domains of the helix and the
beam mode overlap, we observe the same behavior as in
Fig.~\ref{fig3}. Above a certain applied signal amplitude
threshold the distribution function spreads over a much wider
velocity domain, and electrons can be strongly accelerated.

Another striking feature appears in Fig.~\ref{fig4}, which is best
emphasized in the zoom of Fig.~\ref{fig5}a. The transition to
large velocity spread does not occur continuously but rather
occurs by steps when the applied signal amplitude increases. As
shown in Fig.~\ref{fig5}b, plateaus are formed in the measured
velocity distribution function for the maximum interaction length.
A closer look at Fig.~\ref{fig3} also reveals the presence of such
steps in the case of two independently launched waves with equal
amplitudes. We will now show that this generic phenomenon is
related to the intrinsic structure of Hamiltonian phase space for
non integrable systems.

\section{Motion of a test charged particle and Hamiltonian chaos}

We consider charged test particles moving in two electrostatic
waves. The equation modelling the dynamics in this case is
\begin{equation}
\ddot x = \sum_{{\it i}=1}^2 \eta k_i \phi_i\sin(k_i x - \omega_i
t + \varphi_i).
\end{equation}
where $\phi_i$, $k_i$, $\omega_i$ and $\varphi_i$ are respectively
the amplitudes, wave numbers, frequencies and phases of the two
waves; $\eta$ is the charge to mass ratio of the particle.

The motion of the particles governed by this equation is a mixture
of regular and chaotic behaviors mainly depending on the
amplitudes of the waves \cite{Chi,Esc,ElsEsc} and exhibits generic
features of chaotic systems \cite{Lic,Kad,Sch,Ott}. Poincar\'e
sections of the dynamics (a stroboscopic plot of selected
trajectories) are displayed in Fig.~\ref{fig6}. Taking advantage
of the symmetry $x' = -x$, two different cases are superposed in
Fig.~\ref{fig6}. The strength of chaos is measured by the
dimensionless overlap parameter defined as the ratio
\begin{equation}
s =\frac {2(\sqrt{\eta \phi_{1}} + \sqrt{\eta \phi_{2}})}
        {|v_{\varphi 1} - v_{\varphi 2}| }
\end{equation}
of the sum of the resonant velocity half-widths of the two wave
potential wells to their phase velocity difference \cite{Chi}.

For intermediate wave amplitudes, nested regular structures appear
as secondary resonances with a wave number $k_{nm} = nk_1 + mk_2$,
a frequency $\omega_{nm} = n\omega_1 + m\omega_2$, a phase
$\varphi_{nm} = n\varphi_1 + m\varphi_2$ and an effective
amplitude $\phi_{nm} \sim \phi_1^{|n|} \phi_2^{|m|}$ with integer
$n$ and $m$, as shown in the left side of Fig.~\ref{fig6}a for $s
= 0.55$. The self-similar structure of phase space results from
the infinitely nested higher order resonances \cite{DovEsca} which
appear in so-called Arnold tongues between secondary resonances as
predicted by the Poincar\'e-Birkhoff theorem \cite{Arn}.

For larger wave amplitudes, a wide connected zone of chaotic
behavior occurs in between the primary resonances due to the
destruction of the so-called Kolmogorov-Arnold-Moser (KAM) tori
acting as barriers in phase space, as shown in the right side of
Fig.~\ref{fig6}a for $s = 0.85$. Between these two values of $s$,
invariant tori are sequentially destroyed and replaced by cantori.
Transport in velocity (and thereby in $(x,v)$ space) occurs
through the holes of the cantori, which have a fractal structure
\cite{Meiss}. When a torus breaks, the flux through it is still
zero, as the associated turnstile has vanishing area :
trajectories leak easily through its holes only for $s$ values
significantly above its destruction threshold. For the dynamics
$\ddot x = - \varepsilon(\sin(x) +0.16 k \sin k(x-t))$, with  $k =
5/3$, which is close to our experimental conditions, the threshold
for large scale chaos (destruction of the most robust torus) is
near $s\approx 0.75$, as seen from the Poincar\'e section of
orbits over long times (Fig.\ref{fig6}, (b-c)).

If one considers a beam of initially monokinetic particles having
a velocity equal to the phase velocity of one of the waves, this
chaotic zone is associated with a large spread of the velocities
after some time since the particles are moving in the chaotic sea
created by the overlap of the two resonances \cite{Esc,ElsEsc}. In
analogy with the experimental conditions, we compute (with a
second order centered symplectic integrator) the trajectories of 1
million particles initially uniformly spread in position with the
same velocity (equal to 1). As seen in fig.~\ref{fig7}, as the
overlap parameter increases, the beam wiggles more and is spread
over a wider velocity range \cite{EndNote30a}.

As shown in Figs.~\ref{fig8} and $9$, the transition to large
scale chaos occurs by vertical steps, related to the presence of
the nested secondary resonances, with height scaling as the
amplitudes $\phi_{nm}$. Therefore the border of the velocity
domain over which the beam is spread exhibits a devil's
staircase-like behavior. Fig.~\ref{fig8}a is the result of
applying the same treatment as applied to experimental data in
section \ref{SecExp} to the computed velocity distribution
function obtained for different overlap parameters varying from 0
to 1 by steps of $0.01$ keeping the wave amplitude ratio constant
and approximately equal to the experimental value of
Fig.~\ref{fig4}. Each velocity distribution function is obtained
by recording the velocity histogram of our $10^6$ particles over
1600 boxes with normalized velocity width equal to $0.001$ after
an integration time $\tau$ equal to 10 Poincar\'e periods. Even
for an overlap parameter equal to 1, during the integration time
$\tau$, the particles could not explore the whole chaotic sea
between the two main resonances. But the particles could typically
experience one complete trapping motion as shown by the periodic
velocity bunching and related same horizontally wedge-shaped
regions as experimentally observed in Fig.~\ref{fig2}. An
asymmetric plateau \cite{EndNote30b} is formed in the velocity
distribution function as exemplified in the insert of
Fig.~\ref{fig8}a. The detailed analysis of the border of the
distribution function displayed in Fig.~\ref{fig8}b shows that
this plateau can be associated to the secondary resonance
characterized by integers $m = n = 1$.

Fig.~\ref{fig9} is obtained in the same way as Fig.~\ref{fig8} but
corresponds to a longer integration time $\tau$ equal to 20
Poincar\'e periods. Now the particles can explore the whole
chaotic sea in between the two main resonances, and finer details
of the ``devil's staircase'' associated to the presence of the
nested secondary resonances can be observed. In Fig.~\ref{fig9}b
showing the low velocity border of the computed chaotic domain,
the position of some of the most important secondary resonances is
indicated with a label corresponding to their characteristic
integers $(n,m)$. It appears clearly that the presence of a
secondary resonance results in a step in the 2D contour plot of
the distribution function.  The most striking feature is the great
similarity with the above experimental results which can now be
reexamined in the light of the two-waves model.

\section{Discussion}

To allow easier comparison with numerical simulations, we compute
the overlap parameter of the helix and the beam modes. The wave
amplitude can be estimated by fitting the trapping domain with a
parabola as shown in Fig.~\ref{fig2} to Fig.~\ref{fig4}. As
already mentioned, we checked that, for the helix mode, this
estimate is in very good agreement with the amplitude deduced from
measuring the antenna coupling coefficient with three antennas
\cite{Malmberg}. Then we display the upper velocity domain border
(i.e. the location of the upper steep edge of the plateau as shown
in Fig.~\ref{fig5}b) as the applied signal amplitude is varied. We
thus obtain Fig.~\ref{fig5}c. The experimentally observed steps
are clearly related to the presence of secondary resonances
labelled $(1,m)$ with $m$ = 1,~2,~3.

For a different initial beam velocity, Fig.~\ref{fig10}a shows the
upper velocity domain border as the applied signal amplitude is
varied by steps of $1.5$~V; only the steps associated to the main
secondary resonances $(1, n)$ for $n = 1...6$ can be observed. The
points plotted are the outcome of a single set of measurements;
the solid line is a guide to the eye. By reducing the signal
amplitude steps to $0.5$~V, more details of the devil's staircase
can be observed and higher order resonances show up. For example,
in Fig.~\ref{fig10}b, more steps are observed such as $(2,3) =
(1,1) + (1,2)$, $(2,5)$ and $(2,7)$ which stand among the next
strongest resonances in the Farey tree structure. In this latter
plot, two new, independent measurements have been reported (as
circles and squares) \cite{EndNote30c} to give an estimate of the
experimental errors, and the solid line interpolates between their
averages. The behavior reported in Fig.~\ref{fig10}a and
Fig.~\ref{fig10}b is typical of self-similar phenomena.

It is worth mentioning that we focused our attention on the
velocity domain in between the two main resonances. As shown by
the upper part of Fig.~\ref{fig4}a, we also observe vertical steps
associated with resonances (1,-5) and (1,-6); this feature is also
obvious in the numerical simulation of Fig.~\ref{fig9} where a
step associated with resonance (1,-3) clearly appears on the upper
part. Since numerical simulations deal with a restricted number of
particles (typically 1 million), only the steps associated with
lower $m$'s appear.

\section{Conclusion and perspectives}

We have thus exhibited that measuring the energy distribution of
the electron beam at the output of a travelling wave tube allows a
direct experimental exploration of fractal features of the complex
Hamiltonian phase space. It constitutes a direct experimental
evidence of a ``devil's staircase'' in a time-dependent system.

This knowledge opens new tracks in the study of these systems:

-It allowed accurate experiments aimed at investigating
Hamiltonian phase space \cite{APS05} and testing new methods of
control of Hamiltonian chaos \cite{Chan}.

-It is a step in the direction of experimental assessment of the
still open question of quasilinear diffusion in a broad spectrum
of waves \cite{Car} excited with the arbitrary waveform generator
\cite{DovGuy}.

-Time resolved measurements of the evolution of electron bunches
injected with a prescribed phase with respect to a wave \cite{Bou}
should allow to explore more details about the test particle
dynamics.

-The influence of self-consistent effects could be studied by
gradually increasing the beam intensity.

Beside their fundamental importance, these studies can find direct
useful applications in the fields of :

-plasma physics,

-control of complex systems,

-electron devices such as travelling wave tubes and free electron
lasers where improvement of performances are of crucial
importance.

\smallskip
\smallskip
\smallskip

The authors are grateful to J-C.~Chezeaux, D.~Guyomarc'h, and
B.~Squizzaro for their skillful technical assistance, and to
C.~Chandre and D.F.~Escande for fruitful discussions. This work is
supported by Euratom/CEA. A.~Macor benefits from a grant by
Minist\`ere de la Recherche. Y.~Elskens benefited from a
delegation position with Centre National de la Recherche
Scientifique.







\clearpage

\begin{figure}[tbp]
\includegraphics[width=8cm]{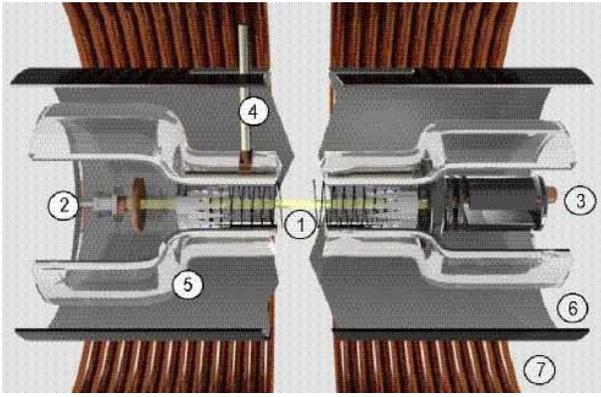}
\caption{ Travelling Wave Tube rendering: (1) helix, (2) electron
gun, (3) trochoidal analyzer, (4) antenna, (5) glass vacuum tube,
(6) slotted rf ground cylinder, (7) magnetic coil. } \label{fig1}
\end{figure}


\begin{figure}[tbp]
\includegraphics[width=8cm]{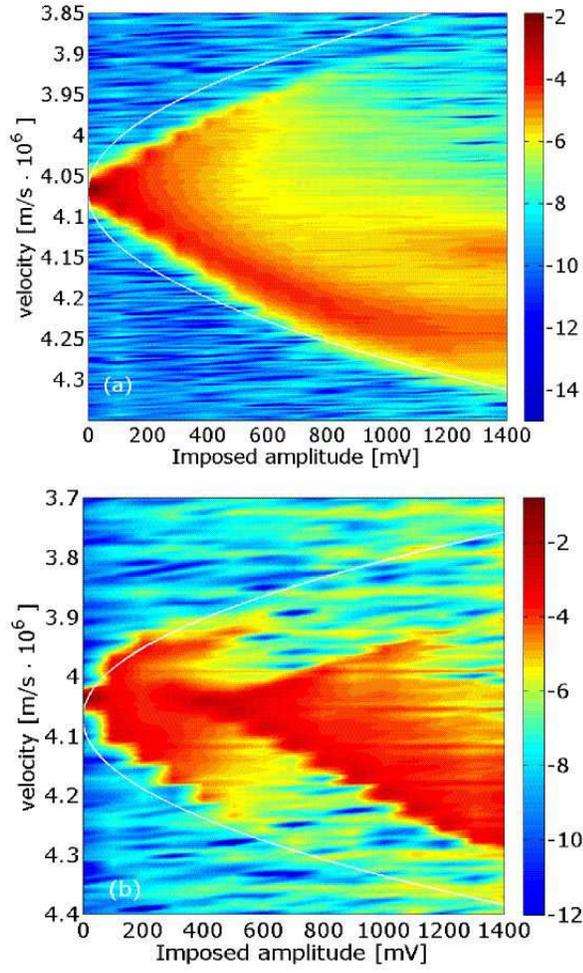}
\caption{\label{fig2} 2D contour plot of the measured velocity
distribution function (with logarithmically scaled color coding)
of a test beam ($I_{\mathrm b} = 120 {\,\rm nA}, v_{\mathrm b} =
4.07 \times10^6 {\,\rm {m/s}}$) trapped in a single wave at 30 MHz
with trapping domain (continuous curve) for increasing amplitude
and fixed interaction length: a) $L = 2.6 {\,\rm m}$, b) $L = 3.5
{\,\rm m}$.}
\end{figure}


\begin{figure}
\includegraphics[width=8cm]{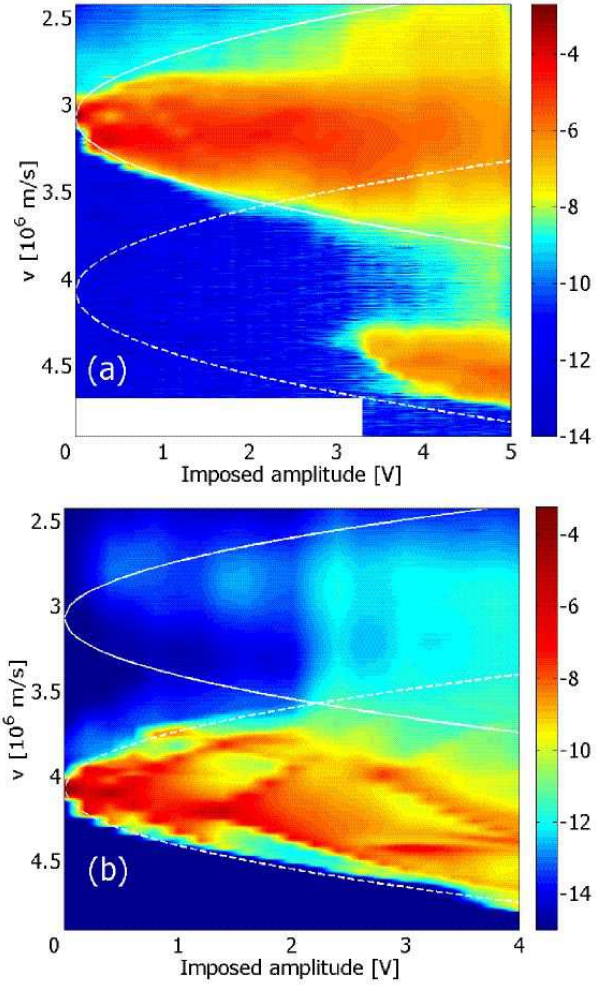}
\caption{2D contour plot of the measured velocity distribution
function (with logarithmically scaled color coding) of a test beam
($I_{\mathrm b} = 10 {\,\rm nA}$) interacting with two waves at 30
MHz and 60 MHz with trapping domains (continuous and dashed
curves) for increasing amplitude and fixed interaction length $L =
3.6 {\,\rm m}$: a) $v_{\mathrm b} = 3.55 \times10^6 {\,\rm
{m/s}}$, b) $v_{\mathrm b} = 4.07 \times10^6 {\,\rm {m/s}}$
\cite{EndNote29}.} \label{fig3}
\end{figure}


\begin{figure}
\includegraphics[width=8cm]{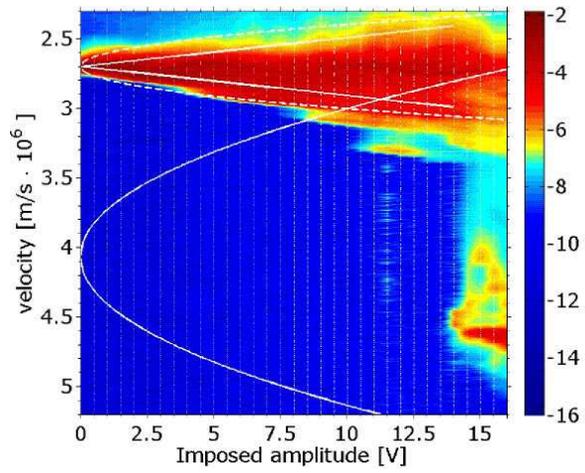}
\caption{\label{fig4} 2D contour plot of the measured velocity
distribution function (with logarithmically scaled color coding)
of a test beam ($I_{\mathrm b} = 10 {\,\rm nA}, v_{\mathrm b} =
2.7 \times10^6 {\,\rm {m/s}}$) with applied signal at 30 MHz for
increasing amplitude and fixed interaction length $L = 3.6 {\,\rm
m}$ (trapping domains of helix and beam modes are indicated by
continuous and dashed parabolas).}
\end{figure}


\begin{figure}
\includegraphics[width=8cm]{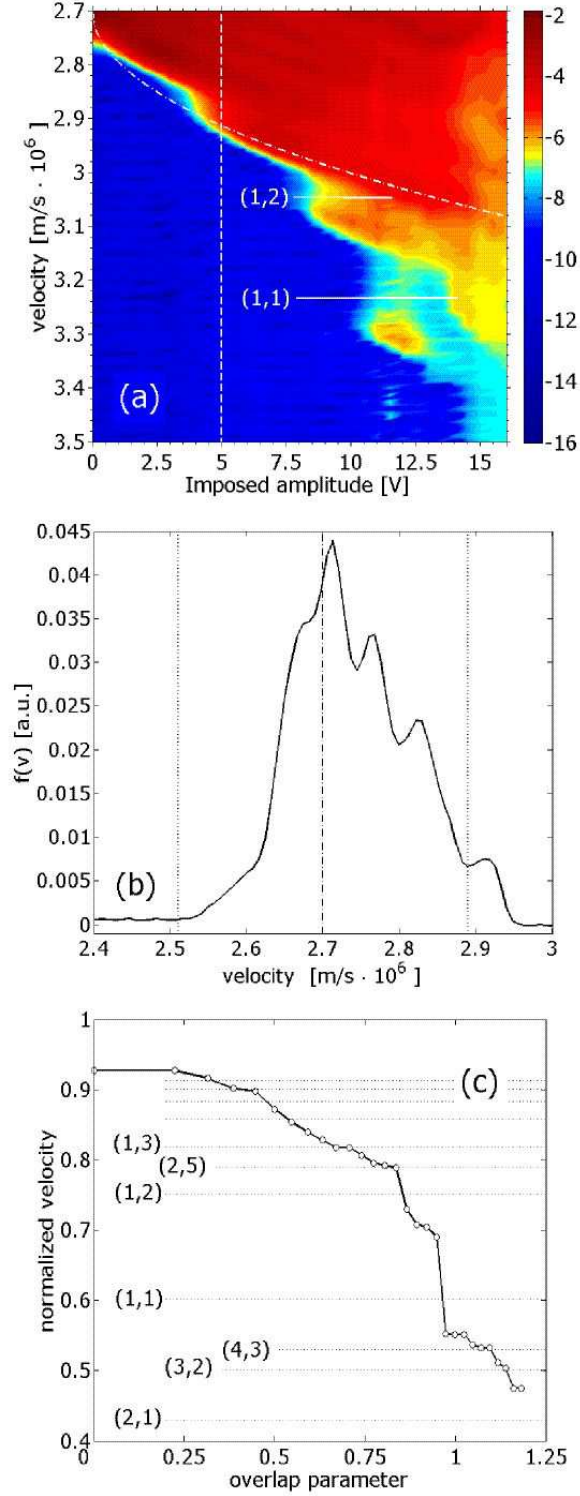}
\caption{\label{fig5} a) Zoom of Fig.~\ref{fig4}. b) Beam velocity
distribution function for amplitude corresponding to the dashed
vertical line in the above figure. c) Normalized upper velocity
frontier of $f(v)$ versus overlap parameter $s$; velocity
normalization is such that $v = 0$ (resp. 1) stands for the helix
(resp. beam) mode phase velocity; secondary resonances $(n,m)$ are
indicated at velocities $m\kappa /(n+m\kappa )$ with $\kappa =
v_{\mathrm \varphi} /v_{\mathrm b}$.}
\end{figure}


\begin{figure}
\includegraphics[width=8cm]{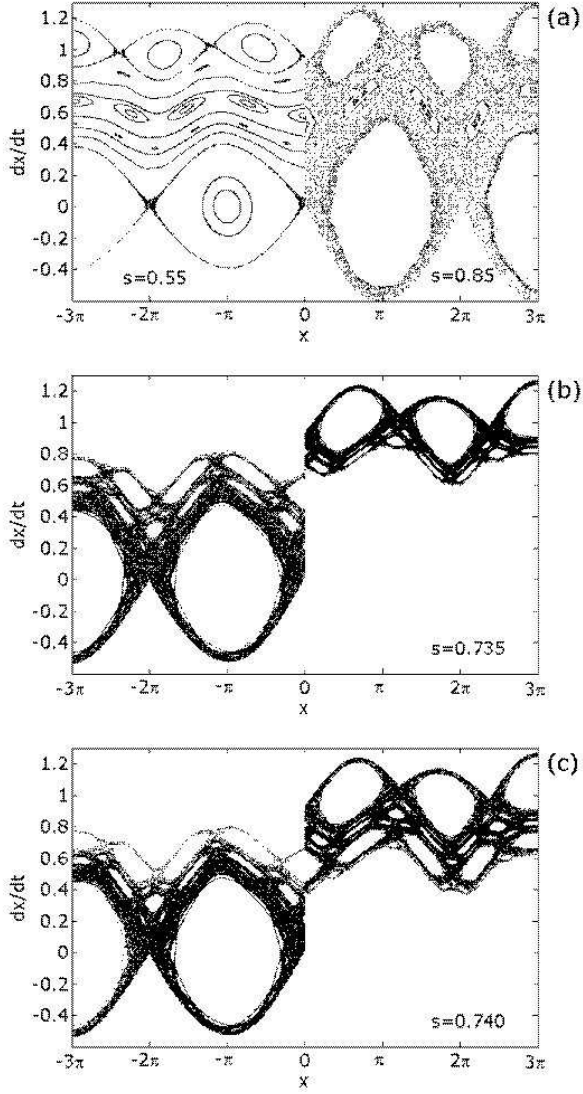}
\caption{Poincar\'e surface of section for the dynamics given by
$\ddot x = - \varepsilon(\sin(x) +0.16 k \sin k(x-t))$, with  $k =
5/3$, for which $s = 2.8\sqrt{\varepsilon}$. (a) Left half for
$\varepsilon \simeq 0.39~(s =0.55)$ exhibits island chains of
secondary resonances at rational velocities $m/(n+m)$, as seen
from 15 orbits which do not mix~; right half for $\varepsilon
\simeq 0.92~(s =0.85)$ exhibits large scale chaos as seen for two
orbits. (b) Two orbits related to upper and lower main chaotic
domains for $s=0.735$, each iterated over $10^5$ Poincar\'e
periods~ ; to allow an easy comparison, only points with $-3\pi <
(x~{\mathrm{mod}}~6 \pi) < 0$ are plotted for one orbit, and $0 <
(x~{\mathrm{mod}}~6 \pi) < 3 \pi$ for the other orbit. (c) Similar
plot for $s=0.74$.} \label{fig6}
\end{figure}


\begin{figure}
\includegraphics[width=8cm]{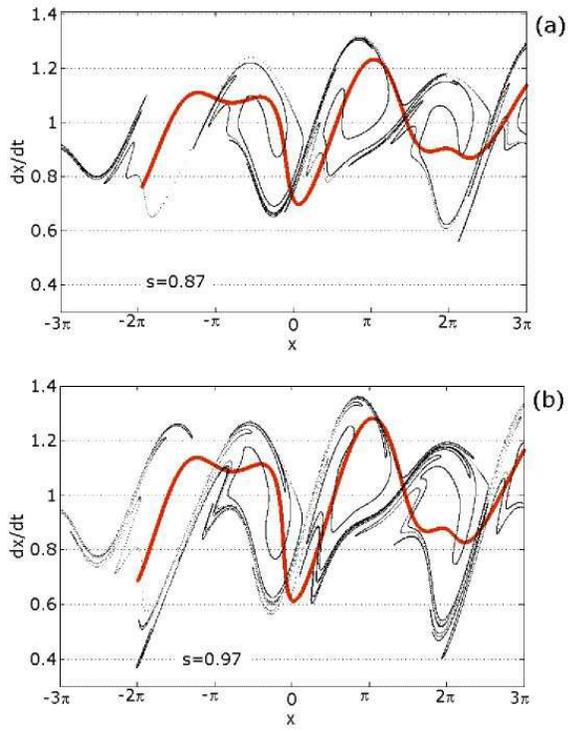}
\caption{Evolution of the beam initially at $v=1$ for the same
dynamics as Fig.~\ref{fig6}, for values of $s$ around the most
prominent step at times $\tau = 6 \pi /5$ (bold, red dots) and
$\tau = 12 \pi$ (small, black dots).} \label{fig7}
\end{figure}


\begin{figure}
\includegraphics[width=8cm]{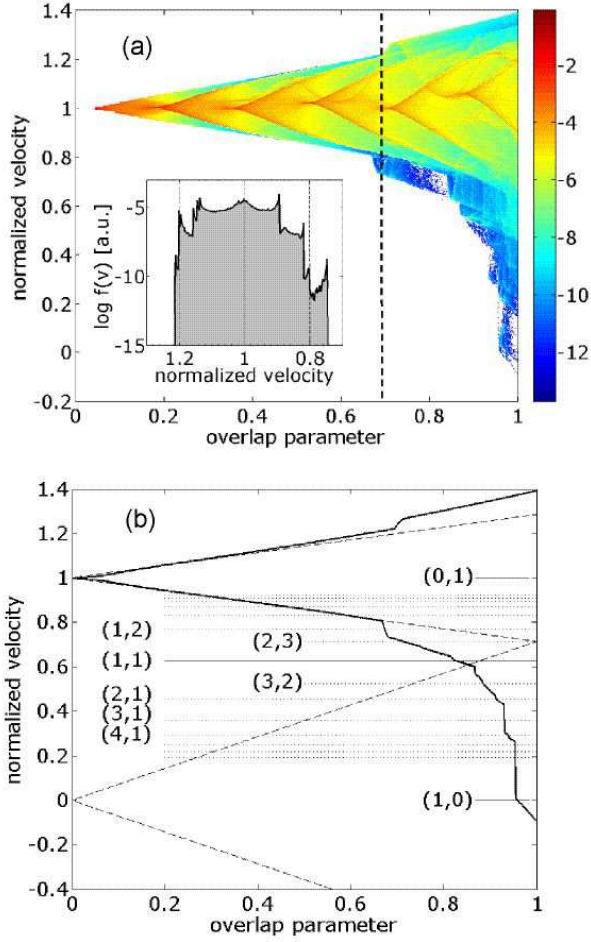}
\caption{Numerical devil's staircase.  a) Probability distribution
function $f(v)$ (with logarithmically scaled color coding) of the
velocity for the dynamics of Fig.~\ref{fig6}, as a function of
overlap parameter $s = 2.8 \sqrt{\varepsilon}$ after a time $\tau
= 12 \pi$, i.e. $10$ Poincar\'e stroboscopic periods (inset,
$f(v)$ for the value of $s$ indicated by the vertical dashed
line). b) Velocity frontiers of $f(v)$ versus $s$ for $t = 12
\pi$; dashed oblique lines starting from $v = 0$ and $v = 1$
indicate the primary resonances trapping domains; secondary
resonances $(n,m)$ are indicated at rational velocities $m k/(n+m
k)$. } \label{fig8}
\end{figure}


\begin{figure}
\includegraphics[width=8cm]{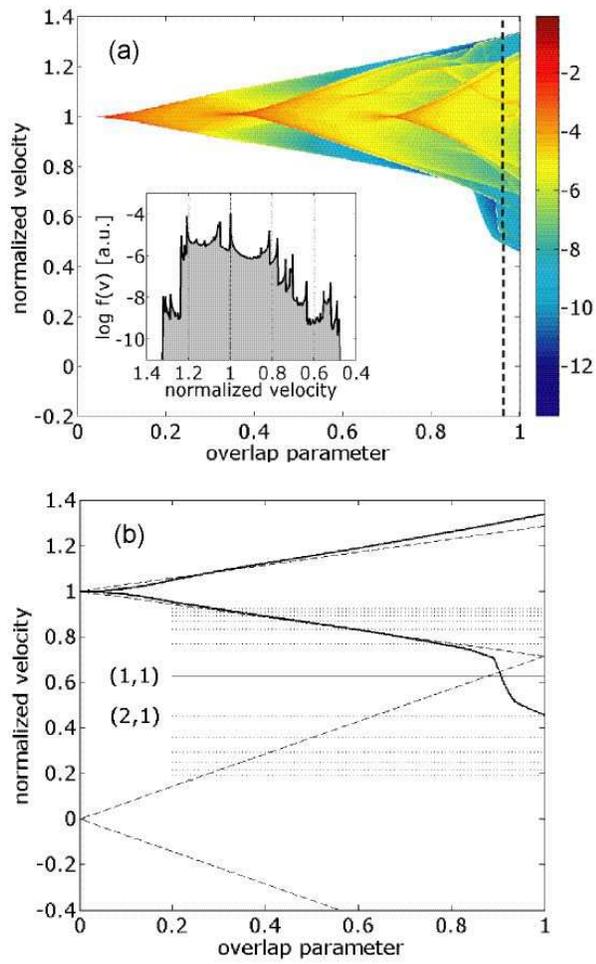}
\caption{Numerical devil's staircase after a time $\tau = 24 \pi$,
i.e. $20$ Poincar\'e stroboscopic periods.  a) Probability
distribution function $f(v)$. b) Velocity frontiers of $f(v)$.}
\label{fig9}
\end{figure}


\begin{figure}
\includegraphics[width=8cm]{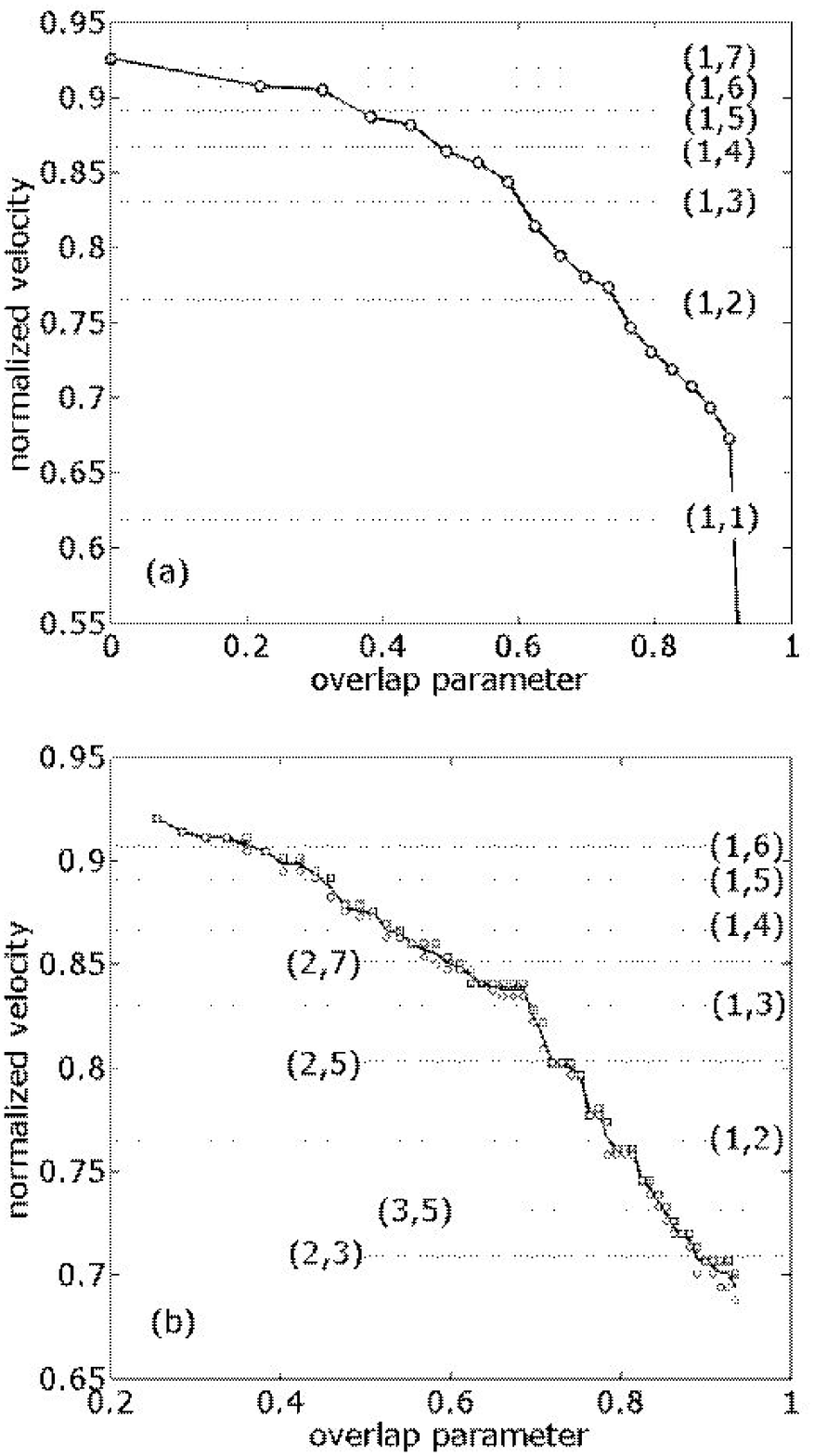}
\caption{Experimental devil's staircase: plot versus overlap
parameter of the upper normalized velocity frontier of $f(v)$ for
a test beam ($I_{\mathrm b} = 10 {\,\rm nA}, v_{\mathrm b} = 2.5
\times10^6 {\,\rm {m/s}}$) with applied signal at 30 MHz for
increasing amplitude and fixed interaction length $L = 3.6 {\,\rm
m}$ with (a) coarse and (b) fine signal amplitude scan; velocity
normalization is such that $v = 0$ (resp. 1) stands for the helix
(resp. beam) mode phase velocity; secondary resonances $(n,m)$ are
indicated at velocities $m\kappa /(n+m\kappa )$ with $\kappa =
v_{\mathrm \varphi} /v_{\mathrm b}$.} \label{fig10}
\end{figure}



\begin{references}

\bibitem{Dim}
{G. Dimonte and J.H. Malmberg, Phys. Fluids {\bf 21}, 1188
(1978).}

\bibitem{Tsu}
{S.I. Tsunoda, F. Doveil and J.H. Malmberg, Phys. Rev. Lett. {\bf
58}, 1112 (1987).}

\bibitem{APS05}
{F.~Doveil and A.~Macor, Phys. Plasmas {\bf 13}, in press (2006).}

\bibitem{DovEsc}
{F. Doveil, D.F. Escande and A. Macor, Phys. Rev. Lett. {\bf 94},
085003 (2005).}

\bibitem{DovAu}
{F. Doveil, Kh. Auhmani, A. Macor and D. Guyomarc'h, Phys. Plasmas
{\bf 12}, 010702 (2005).}

\bibitem{Chi}
{B.V. Chirikov, Phys. Rep. {\bf 52}, 263 (1979).}

\bibitem{Esc}
{D.F. Escande, Phys. Rep. {\bf 1212}, 165 (1985).}

\bibitem{Cha}
{C. Chandre and H.R. Jauslin, Phys. Rep. {\bf 365}, 1 (2002).}

\bibitem{Ko}
{H. Koch, Disc. Cont. Dyn. Syst. {\bf 11}, 881 (2004).}

\bibitem{ElsEsc}
{Y. Elskens and D.F. Escande, {\it Microscopic dynamics of plasmas
and chaos} (IoP Publishing, Bristol, 2003).}

\bibitem{DovEsca}
{F. Doveil and D.F. Escande, Phys. Lett. {\bf 90A}, 226 (1982).}

\bibitem{Lic}
{A.J. Lichtenberg and M.A. Lieberman, {\it Regular and chaotic
dynamics} (Springer, New York, 1992).}

\bibitem{Kad}
{L.P. Kadanoff, {\it From order to chaos: Essay: Critical, chaotic
and otherwise} (World Scientific, Singapore, 1993); {\it II}
(1998).}

\bibitem{Sch}
{H.G. Schuster, {\it Deterministic chaos} (VCH Verlag, Weinheim,
1988).}

\bibitem{Ott}
{E. Ott, {\it Chaos in dynamical system} (Cambridge university
press, Cambridge, 1993).}

\bibitem{Gor}
{L.P. Gor'kov and G. Gr\"{u}ner, {\it Charge density waves in
solids} (Elsevier, Amsterdam, 1989).}

\bibitem{Ori}
{S.E. Hebboul and J.C. Garland, Phys. Rev {\it B} {\bf 47}, 5190
(1993); E. Orignac and T. Giamarchi, Phys. Rev. {\it B} {\bf 64},
144515 (2001).}

\bibitem{Flo}
{L. Floria and F. Falo, Phys. Rev. Lett.  {\bf 68}, 2713 (1992);
O.M. Braun, A.R. Bishop, and J. R\"{o}der, Phys. Rev. Lett. {\bf
79}, 3692 (1997).}

\bibitem{Hup}
{P. Pieranski, P. Sotta, D. Rohe, and M. Imperor-Clerc, Phys. Rev.
Lett. {\bf 84}, 2409 (2000); M. Hupalo, J. Schmalian and M.C.
Tringides, Phys. Rev. Lett. {\bf 90}, 216106 (2003).}

\bibitem{Ade}
{D. Adelman, C.P. Burmester, L.T. Wille, P.A. Sterne and R.
Gronsky, J. Phys.: Cond. Matter {\bf 4}, L585 (1992).}

\bibitem{Das}
{S. Das Sarma and A. Pinczuk, {\it Perspective in quantum Hall
effects} (Wiley, New York, 1997).}

\bibitem{Gop}
{A. Gopinathan and D.G. Grier, Phys. Rev. Lett. {\bf 92}, 130602
(2004).}

\bibitem{Mac}
{A. Macor, F. Doveil, and Y. Elskens, Phys. Rev. Lett. {\bf 95},
264102 (2005).}

\bibitem{Pie}
{J.R. Pierce, {\it Traveling wave tubes} (Van Nostrand, New York,
1950).}

\bibitem{Gil}
{A.S. Gilmour Jr, {\it Principles of traveling wave tube} (Artech
House, London, 1994).}

\bibitem{Guy}
{D. Guyomarc'h and F. Doveil, Rev. Sci. Instrum. {\bf 71}, 4087
(2000).}

\bibitem{Malmberg}
{J.H. Malmberg, T.H. Jensen and T.M. O'Neil, Plasma Phys.
Controlled Nucl. Fusion Res. {\bf 1}, 683 (1966)}

\bibitem{Myn}
{H.E. Mynick and A.N. Kaufman, Phys. Fluids {\bf 21}, 653 (1978).}

\bibitem{EndNote29} {Two causes can contribute to the presence of
slower electrons: i) the gun itself, ii) space charge effects in
the trochoidal analyzer. The normalization of the velocity
distribution function enhances this lower tail when the beam is
more spread (hence the observed periodicity at low wave amplitude,
low electron velocity in Fig.~\ref{fig3}b). Fig.~\ref{fig3}a which
focuses on accelerated electrons is not polluted by these low
energy electrons in the relevant domain.}

\bibitem{Arn}
{V.I. Arnold, {\it Mathematical methods of classical mechanics}
(Nauka, Moscow, 1974).}

\bibitem{Meiss} J.D. Meiss, Rev. Mod. Phys. {\bf 64}, 795 (1992).

\bibitem{EndNote30a} {As the beam intersects the core of the cat eye,
part of it merely rotates in the trapping domain. One should not
mistake the image of the beam at later times, as in
Fig.~\ref{fig7}, with the unstable manifold of the hyperbolic
point associated with the wave. However, the wiggles of the beam
in the chaotic domain must follow grossly the wiggles of this
unstable manifold.}

\bibitem{EndNote30b} {The many small peaks in the plot of $f(v)$
are the result of the foldings of the smooth curve in $(x,v)$
space which was initially the monokinetic beam at $v=1$. The
probability density in $v$ diverges at each fold like the
classical arc sine law.}

\bibitem{EndNote30c} {The squares are above the circles because
they correspond to a lower estimate of the noise level.}

\bibitem{Chan}
{C. Chandre, G.~Ciraolo, F.~Doveil, R.~Lima, A.~Macor and M.~
Vittot, Phys. Rev. Lett {\bf 94}, 074101 (2005).}

\bibitem{Car}
{J.R. Cary, D.F. Escande, and A.D. Verga, Phys. Rev. Lett. {\bf
65}, 3132 (1990).}

\bibitem{DovGuy}
{F. Doveil and D. Guyomarc'h, Comm. Nonlinear Sci. Numerical
Simulation {\bf 8}, 529 (2003).}

\bibitem{Bou}
{A. Bouchoule and M. Weinfeld, Phys. Rev. Lett. {\bf 36}, 1144
(1976).}

\end{references}
\end{document}